# Deep Neural Networks for Automated Classification of Colorectal Polyps on Histopathology Slides: A Multi-Institutional Evaluation

Jason W. Wei[1,2]; Arief A. Suriawinata, MD[3]; Louis J. Vaickus, MD, PhD[3]; Bing Ren, MD, PhD[3]; Xiaoying Liu, MD[3]; Mikhail Lisovsky, MD, PhD[3]; Naofumi Tomita, MS[1]; Behnaz Abdollahi, PhD[1]; Adam S. Kim, MD[4]; Dale C. Snover, MD[5]; John A. Baron, MD[6]; Elizabeth L. Barry, PhD[7]; Saeed Hassanpour, PhD[1,2,7*]

[1]Department of Biomedical Data Science, Dartmouth College, Hanover, NH, USA

[2]Department of Computer Science, Dartmouth College, Hanover, NH, USA

[3]Department of Pathology and Laboratory Medicine, Dartmouth-Hitchcock Medical Center, Lebanon, NH, USA

[4]Minnesota Gastroenterology PA, Minneapolis, MN, USA

[5]Department of Pathology, Fairview Southdale Hospital, Edina, MN, USA

[6]Department of Medicine, University of North Carolina, Chapel Hill, NC, USA

[7]Department of Epidemiology, Dartmouth College, Hanover, NH, USA

* Corresponding Author: Saeed Hassanpour, PhD

Postal address: One Medical Center Drive, HB 7261, Lebanon, NH 03756, USA

Telephone: (603) 650-1983

Email: Saeed.Hassanpour@dartmouth.edu

Word Count: 2,968

# KEY POINTS

**Question**: Are deep neural networks trained on data from a single institution for classification of colorectal polyps on digitized histopathology slides generalizable across multiple external institutions?

**Findings**: A new deep neural network was developed based on 326 slide images from our institution to classify the four most common polyp types on digitized histopathology slides. In addition to evaluation on an internal test set of 157 slide images, we evaluated the model on an external test set of 238 slide images from 24 institutions across 13 states in the United States. This model achieved mean accuracies of 93.5% and 87.0% on the internal and external test sets, respectively, which were comparable with the performance of local pathologists on these test sets.

**Meaning**: Deep neural networks could provide a generalizable approach for the classification of colorectal polyps on digitized histopathology slides and if confirmed in clinical trials, could potentially improve the efficiency, reproducibility, and accuracy of one of the most common cancer screening procedures.

# ABSTRACT

**Importance**: Histological classification of colorectal polyps plays a critical role in both screening for colorectal cancer and care of affected patients. An accurate and automated algorithm for the classification of colorectal polyps on digitized histopathology slides could benefit clinicians and patients.

**Objective**: Evaluate the performance and assess the generalizability of a deep neural network for colorectal polyp classification on histopathology slide images using a multi-institutional dataset.

**Design, Setting, And Participants**: In this study, we developed a deep neural network for classification of four major colorectal polyp types, tubular adenoma, tubulovillous/villous adenoma, hyperplastic polyp, and sessile serrated adenoma, based on digitized histopathology slides from our institution, Dartmouth-Hitchcock Medical Center (DHMC), in New Hampshire.

**Main Outcomes and Measures**: We evaluated the deep neural network on an internal dataset of 157 histopathology slide images from DHMC, as well as on an external dataset of 238 histopathology slide images from 24 different institutions spanning 13 states in the United States. We measured accuracy, sensitivity, and specificity of our model in this evaluation and compared its performance to local pathologists' diagnoses at the point-of-care retrieved from corresponding pathology laboratories.

**Results**: For the internal evaluation, the deep neural network had a mean accuracy of 93.5% (95% CI 89.6%–97.4%), compared with local pathologists' accuracy of 91.4% (95% CI 87.0%–95.8%). On the external test set, the deep neural network achieved an accuracy of 87.0% (95%

CI 82.7%–91.3%), comparable with local pathologists' accuracy of 86.6% (95% CI 82.3%–90.9%).

**Conclusions and Relevance**: If confirmed in clinical settings, our model could assist pathologists by improving the diagnostic efficiency, reproducibility, and accuracy of colorectal cancer screenings.

# INTRODUCTION

In the United States, colorectal cancer is estimated to cause 51,020 deaths in 2019, making it the second-most common cause of cancer death.[1] This death rate, however, has been dropping in the past several decades likely due to successful cancer screening programs.[2-5] Colonoscopy is the most common test in these screening programs in the U.S.[6] During colonoscopies, clinicians excise colorectal polyps and visually examine them on histopathology slides for neoplasia, ultimately reducing the mortality rate by detecting cancer at an early, curable stage and removing preinvasive adenomas or serrated lesions.[7-9] Furthermore, the numbers and types of polyps detected can indicate future risk for malignancies and are therefore used as the basis for subsequent screening recommendations.[6] An algorithm for automated classification of colorectal polyps could potentially benefit cancer screening programs by improving efficiency, reproducibility, and accuracy, as well as reducing the access barrier to pathological services.[10]

In recent years, a class of computational models known as deep neural networks has driven substantial advances in the field of artificial intelligence. Comprising many processing layers, deep neural networks take a data-driven approach to automatically learn the most relevant features of input data for a given task, dramatically improving the state-of the-art in computer vision,[11] natural language processing,[12] and speech recognition.[13] For medical image analysis in particular, deep learning has achieved considerable performance in classification of images including chest radiographs,[14] retinal fundus photographs,[15] head CT scans,[16] lung histopathology slides,[17] and skin cancer images.[18]

In this study, we used 326 slides from our institution to train a deep neural network for colorectal polyp classification. In addition to the standard internal evaluation on 157 slides, we tested our algorithm on an external test set of 238 slides from 24 institutions in the United States and compared our algorithm's performance to that of local pathologists in terms of accuracy, sensitivity, and specificity. To the best of our knowledge, this study is the first to comprehensively evaluate a deep learning algorithm for colorectal polyp classification and show the generalizability of this model across multiple institutions.

# MATERIALS AND METHODS

**Data Collection.** In this study, we utilized an internal and an external dataset of hematoxylin and eosin (H&E) stained formalin-fixed paraffin embedded colorectal polyp whole-slide images. Each of these slides could contain one or more tissue section/polyp.

Our internal dataset was collected from January 2016 to June 2016 at Dartmouth-Hitchcock Medical Center (DHMC), a tertiary academic care center in New Hampshire, USA. This dataset included 508 slides from four most common polyp types, tubular adenoma, tubulovillous/villous adenoma, hyperplastic polyp, and sessile serrated adenoma, according local diagnoses parsed from pathology reports. The slides were scanned using Leica Aperio AT2 scanners at 40x resolution (0.25 µm pixel$^{-1}$) at DHMC. In this internal dataset, each whole-slide image belonged to a different patient and colonoscopy procedure. We partitioned these slides into a training set of 326 slides, a validation set of 25 slides, and an internal test set of 157 slides. The distribution of polyp types was balanced in the validation and internal test sets, while slides were oversampled for hyperplastic polyps and sessile serrated adenomas in the training set to improve model training for these classes (**Figure 1**).

For our external dataset, we collaborated with investigators from a randomized clinical trial on the effect of supplementation with calcium and/or vitamin D for the prevention of colorectal adenomas[21] and their network of laboratories for additional evaluation. Through this collaboration, we were given access to 1,182 whole-slide images along with their diagnoses given by local pathologists. These slides were borrowed from various pathology laboratories in the United States (see **Supplementary Table 1**) by our co-investigator (ELB) from January 2016 to December 2017, digitized by Leica Aperio AT2 scanners at 40x resolution at DHMC (similar to the internal dataset), before returning to the original laboratories. Of these slides, we randomly sampled up to 95 slides for each of four polyp types as diagnosed by the local pathologist. Of note, fifteen of these randomly selected slides were removed due to poor slide quality as determined by our study's lead expert pathologist (AAS). In total, our final external validation set comprised 238 slides from 24 different institutions spanning 13 states in the United States. In this external test set, some of the slides corresponded to the same patients, as our 238 slides came from 179 distinct patients. All slides from the internal and external test sets were held out from model development until final evaluation of the model.

**Data Annotation.** Our study's annotation process involved five gastrointestinal (GI) pathologists from the Department of Pathology and Laboratory Medicine at DHMC: three with GI pathology fellowship training and two who gained GI pathology expertise through years on GI pathology service. For 157 whole slide images in the training set, two of the GI pathologists (AAS and LJV) identified the polyps on the slides and manually annotated a rectangular bounding box around the polyps and normal tissue using Rectlabel[22] annotation tool as the regions of interest for model training. In total, 3,848 regions of interest were identified and labeled as one of the four polyp classes.

We also collected a smaller collection of annotations from 25 separate whole-slide images as our validation set for hyperparameter tuning of our model. In this validation set, the same two pathologists annotated non-overlapping patches of 224 by 224 pixels (or 448 by 448 μm) of classic examples for each polyp type. Since this dataset was used for cross-validation, all fixed-size patches were confirmed with high confidence by both pathologists, and patches with disagreements were discarded.

For the internal test set, our five GI pathologists independently and retrospectively diagnosed each slide as one of the four polyp types. For this internal set, the local diagnoses given at DHMC may have been from one of the five study GI pathologists, but the original diagnosis and identity of the pathologist at the point-of-care were hidden during the retrospective annotation phase.

For the external test set, the five GI pathologists from DHMC also retrospectively diagnosed all slides in the test set in the same fashion as for the internal test set. In total, we had five complete sets of diagnoses from GI pathologists, as well as the diagnoses given by local pathologists at the point-of-care. For both internal and external test sets, gold-standard diagnoses were assigned by taking the majority vote of our five GI pathologists. **Figure 1** depicts the data flow for our study design. **Supplementary Figure 1** shows the number of patches per slide and slide sizes for the internal and external test sets.

**Deep Learning Model.** In this study, we implemented the deep residual network (ResNet), a neural network architecture that significantly outperformed all other models on the ImageNet and COCO image recognition benchmarks.[21] For model training, we applied a sliding window method to the 3,848 variable-size regions of interest labeled by pathologists in the training set, extracting approximately 7,000 fixed-size 224 by 224 pixels patches per polyp type. Then, we initialized ResNet with the MSRA weight initialization[11] and trained our neural network for 200 epochs with an initial learning rate of 0.001, which decayed by a factor of 0.9 every epoch. Throughout training, we applied standard image augmentation techniques including rotations and flips, as well as color jittering on the brightness, contrast, saturation, and hue of each image. For our final model, we used an ensembled model comprising five ResNets of 18, 34, 50, 101, and 152 layers. Overall, training these networks took about 96 hours using a single NVIDIA Tesla K40c graphics processing unit (GPU).

**Slide-Level Inference.** For our deep learning model to infer the overall diagnosis of a whole-slide image, we designed a hierarchical classification algorithm to match the nature of our classification task. First, each slide was broken down into many patches using a sliding window algorithm, and each patch was classified by the neural network.

Using the predicted diagnoses by the neural network for all patches in a given slide, our model first determined whether a polyp was adenomatous (tubular/tubulovillous/villous) or serrated (hyperplastic/sessile serrated) by comparing the number of predicted patches for the adenomatous and serrated types. Adenomatous polyps with more than a certain amount of tubulovillous/villous tissue (>30%) were classified as overall tubulovillous/villous adenoma, whereas the remaining polyps were classified as tubular adenoma. For serrated polyps, our algorithm classified polyps with above a certain amount of sessile serrated patches (>1.5%) as overall sessile serrated adenomas and the remaining polyps as hyperplastic. All thresholds were determined using a grid search over our internal training set and are shown in **Figure 2**. The hierarchical nature of our inference heuristic allowed us to imitate the schema used by pathologists for this classification task without training a separate machine learning classifier.

**Evaluation.** For final evaluation, we compared the performance of our model with that of local pathologists originally made at the point-of-care on both the internal test set and the multi-institutional external test set. Local pathologist performance measures were averaged over all samples, since information about individual pathologists' performances were anonymized. To assess the quality of annotations in our study, we measured the agreement of our GI pathologists in terms of multi-class Cohen's κ. The application of our final model on a whole-slide image in the test sets took less than sixty seconds on average using a single NVIDIA Tesla K40c GPU. For our model's classifications, we calculated accuracy, sensitivity, and specificity in relation to gold-standard diagnoses and compared these metrics with those of local pathologists. Furthermore, we calculated confusion matrices for local pathologists and our model and conducted appropriate error analysis.

# RESULTS

**Table 1** shows the per-class and mean performance metrics of local pathologists and our proposed model for both internal and external test sets.

**Internal evaluation.** For the internal test set from DHMC, inter-observer agreement, measured by Cohen's κ, was in the substantial range (0.61–0.80), with the five study GI pathologists achieving average multi-class Cohen's κ of 0.72 (95% CI 0.64–0.80). Our model achieved a mean accuracy (the unweighted average of individual polyp type accuracies) of 93.5% (95% CI 89.6%–97.4%) compared with local pathologists' accuracy of 91.4% (95% CI 87.1%–95.8%) on the internal dataset. A two-tailed t-test for proportions revealed, however, that the differences in performance were not significant, with *p*-values of 0.48, 0.14, and 0.80, for accuracy, sensitivity, and specificity, respectively.

**Multi-institutional external evaluation.** Our external dataset, on the other hand, had slightly less agreement for both pathologists and our model. Here, the five study GI pathologists achieved an average multi-class Cohen's κ of 0.67 (95% CI 0.60–0.75). With an accuracy of 87.0% (95% CI 82.7%–91.3%) on the external test set, we found that our model performed at a similar level of accuracy, sensitivity, and specificity as local pathologists on this dataset, with p-values of 0.90, 0.86, and 0.99, respectively. Table 1 shows the performance metrics for both local pathologists and deep neural network for each polyp class on both the internal and external test sets.

**Confusion matrices and error analysis.** Moreover, in **Figure 3**, we calculated confusion matrices for both local pathologists and our model on the external test set to determine which polyp types were the most challenging to diagnose. Local pathologists often classified tubular adenomas as tubulovillous/villous adenomas (46%) and hyperplastic polyps as sessile serrated adenomas (14%). The deep neural network similarly classified many tubular adenomas as tubulovillous/villous adenomas (23%) and hyperplastic polyps as sessile serrated adenomas (28%). For further analysis of our model's errors, **Supplementary Figure 2** shows violin plots for predicted percentage areas of each polyp type on slides.

**Visualization.** Finally, we visualized the results of our model on digitized slides by highlighting the regions that contributed to the whole-slide classification. **Figure 4** shows examples of slides with our lead GI pathologist (AAS) annotations, the heatmap detected by our model, and the visualization of our model's results.

# DISCUSSION

To the best of our knowledge, this study is the first to evaluate a deep neural network for colorectal polyp classification on a large multi-institutional dataset with comparison to local diagnoses made at the point-of-care. On a test set comprising 238 images from 24 external institutions, our model achieved an accuracy of 87.0%, which was on par with the local pathologists' accuracy of 86.6% at the $\alpha=0.05$ level. Regarding annotation agreement, the five study GI pathologist annotators had an average Cohen's κ of 0.72 (95% CI 0.64-0.80) on our internal test set and 0.67 (95% CI 0.60-0.75) on the external test set, much higher than previously reported Cohen's κ scores of 0.46,[22] 0.31,[23] 0.55,[24] and 0.54.[25] This difference in performance is likely due to differences in polyp type distributions in various datasets, inter-laboratory variations in tissue processing and staining, and institutional biases in the polyp classification criteria.

In terms of error analysis, the deep neural network made the same types of misclassifications as local pathologists, as shown by the similarities in their confusion matrices. Both the model and local pathologists distinguished between adenomatous (tubular/tubulovillous/villous) and serrated (hyperplastic/sessile serrated) polyps with high accuracy but found further sub-classification more challenging. We hypothesize that many of the mistakes in sub-classification occurred because thresholds for detection of tubulovillous/villous growths and of sessile serrated crypts vary among pathologists, as our lead GI pathologist's manual inspection of discordances found that many of the errors made by the deep neural network were similar to mistakes made by pathologists in practice. Furthermore, subjective examination of the visualization results by our study's lead GI pathologist confirmed that the highlighted regions of interest are mostly on target.

Our study not only demonstrates the utility of a deep learning model for classification of colorectal polyps but also advances previous literature in terms of model evaluation and study design. The previous foremost study on deep learning for colorectal polyp classification, done by our team,[26,27] demonstrated good performance on an internal dataset but used a simpler approach and did not include pathologist-level performance or local diagnoses. Our study, on the other hand, evaluates a deep neural network on a multi-institutional external dataset and demonstrates a comparable diagnostic performance of deep neural networks compared to local pathologists at the point-of-care. Many previous papers demonstrated clinician-level performance of deep neural networks on various medical classification tasks.[14-18, 28-29] All of these studies, however, measured clinician-level performance on a predetermined number of clinicians from a few medical institutions in a controlled setting. Although it is significant to measure retrospective clinician performance on classification tasks, we used diagnoses by local pathologists in clinical practice at the point-of-care in 24 external institutions for comparison against our deep neural network.

A deep learning model for colorectal polyp classification, if validated through clinical trials, has potential for widespread application in clinical settings. Our model could be implemented in laboratory information systems to guide pathologists by identifying areas of interest on digitized slides, which could improve work efficiency, reproducibility, and accuracy for colorectal polyp classification. Although expert clinician confirmation of diagnoses will still be required, our model could triage slides with diagnoses that are more likely to be preinvasive for subsequent review by pathologists. Since the U.S. Preventive Services Task Force (USPSTF) recommends that all adults aged 50 to 75 undergo screening for colorectal cancer, an automated model for classification could be useful in relieving pathologists' burden in slide review and ultimately reduce the barrier of access for colorectal cancer screening.

Our study has several limitations. Although our model performed on par with local pathologists on the external test set, it did not perform as well as the internal evaluation, as shown in **Table 1**. Our results suggest that there is a higher level of variability among slides from various institutions and our model can be further improved by training on larger diverse datasets. Furthermore, although our model identifies the most common polyp types, our study was done on well-sectioned, clearly stained slides and did not include less-common classes such as traditional serrated adenoma or sessile serrated adenoma with cytological dysplasia. Also, our model was not evaluated on entirely normal slides. Our team plans to collect further data and extend our model and its evaluation to these additional cases as future work. Finally, local pathologists might have had access to additional slides and patient information, such as patient colonoscopy history and polyp biopsy location, that influenced their diagnoses of polyp type. Access to this additional information might explain some of the discrepancies between local diagnoses and the ground-truth labels, which were only based on digitized slides.

Moving forward, further work can be done in deep learning for analysis of colorectal polyp images. Foremost, we plan to implement our model prospectively in a clinical setting to measure its ability to enhance pathologists' classification of colorectal polyps and improve outcomes in a clinical trial. In terms of technical improvements to our model, more data can be collected and used for training to increase the model's performance, especially for sessile serrated adenomas. Moreover, related work has shown that deep learning has potential to identify hidden features in histopathology images that can be used to detect gene mutations[19] and predict patient survival,[30-32] tasks that pathologists do not perform. To this end, we plan to collect more patient outcome data to train our model to predict polyp recurrence and patient survival in colorectal cancer.

## CONCLUSIONS

In this study, we developed a deep neural network model to classify the four most common colorectal polyp types. Our model was evaluated on internal and external test sets. We also comprehensively compared the performance of this model to local pathologists' diagnoses at the point-of-care retrieved from corresponding pathology labs. Our evaluation on the internal and external test sets showed that the performance of our model was on par with the performance of local pathologists. If confirmed in clinical trials, our model could potentially improve the efficiency, reproducibility, and accuracy of one of the most common cancer screening procedures.

# ACKNOWLEDGEMENTS

The authors would like to thank Thomas H. Cormen, PhD and Lamar Moss for their feedback on this paper and Leila Mott for her help with the dataset. We also thank Minnesota Gastroenterology for their help with data collection.

# FUNDING

This work is supported by grants from the National Institutes of Health (R01LM012837, P20GM104416, R01CA098286), Geisel School of Medicine at Dartmouth, and the Norris Cotton Cancer Center.

# ROLE OF THE FUNDER/SPONSOR

The funding sources had no role in the design and conduct of the study; collection, management, analysis, and interpretation of the data; preparation, review, or approval of the manuscript; and decision to submit the manuscript for publication.

# COMPETING INTERESTS

There authors declare no competing interests.

# HUMAN SUBJECT REGULATIONS

This study and the use of human subject data in this project were approved by the Dartmouth-Hitchcock Health institutional review board (D-HH IRB) with a waiver of informed consent. The conducted research reported in this paper is in accordance with this approved D-HH IRB protocol and the World Medical Association Declaration of Helsinki on Ethical Principles for Medical Research Involving Human Subjects.

# SOFTWARE AND CODE AVAILABILITY

The algorithms in this study were implemented in Python (version 3.6). We used OpenSlide (version 3.4.1) to convert the digitized image format and PyTorch (version 0.4) for training the deep neural network models. The statistical analysis and confidence intervals were calculated using the Statistics (version 3.4) library in python. The source code for this study is publicly available at https://github.com/BMIRDS/deepslide.

# TABLES

**Table 1.** Per-class comparison between local pathologists and our deep neural network in classifying colorectal polyps on an internal test set of 157 slides and an external test set of 238 slides. TA: tubular adenoma; TVA: tubulovillous/villous adenoma; HP: hyperplastic polyp; SSA: sessile serrated adenoma; Acc: accuracy; Sens: sensitivity; Spec: specificity.

| Polyp Type | Internal Test Set (n=157) | | | | | | External Test Set (n=238) | | | | | |
|---|---|---|---|---|---|---|---|---|---|---|---|---|
| | Local Pathologists | | | Deep Neural Network | | | Local Pathologists | | | Deep Neural Network | | |
| | Acc | Sens (%) | Spec | Acc | Sens (%) | Spec | Acc | Sens (%) | Spec | Acc | Sens (%) | Spec |
| TA | 89.8 | 76.1 | 95.5 | 93.0 | 89.1 | 94.6 | 79.8 | 53.7 | 97.2 | 84.5 | 73.7 | 91.6 |
| TVA | 94.3 | 88.2 | 95.8 | 95.5 | 97.1 | 95.1 | 81.5 | 100.0 | 77.7 | 89.5 | 97.6 | 87.8 |
| HP | 89.8 | 76.9 | 94.1 | 92.4 | 82.1 | 95.8 | 91.6 | 80.8 | 96.8 | 85.3 | 60.3 | 97.5 |
| SSA | 91.7 | 81.6 | 95.0 | 93.0 | 78.9 | 97.5 | 93.3 | 79.2 | 94.8 | 88.7 | 79.2 | 89.7 |
| Mean | 91.4 | 80.7 | 95.1 | 93.5 | 86.8 | 95.7 | 86.6 | 78.4 | 91.6 | 87.0 | 77.7 | 91.6 |

**FIGURE LEGENDS**

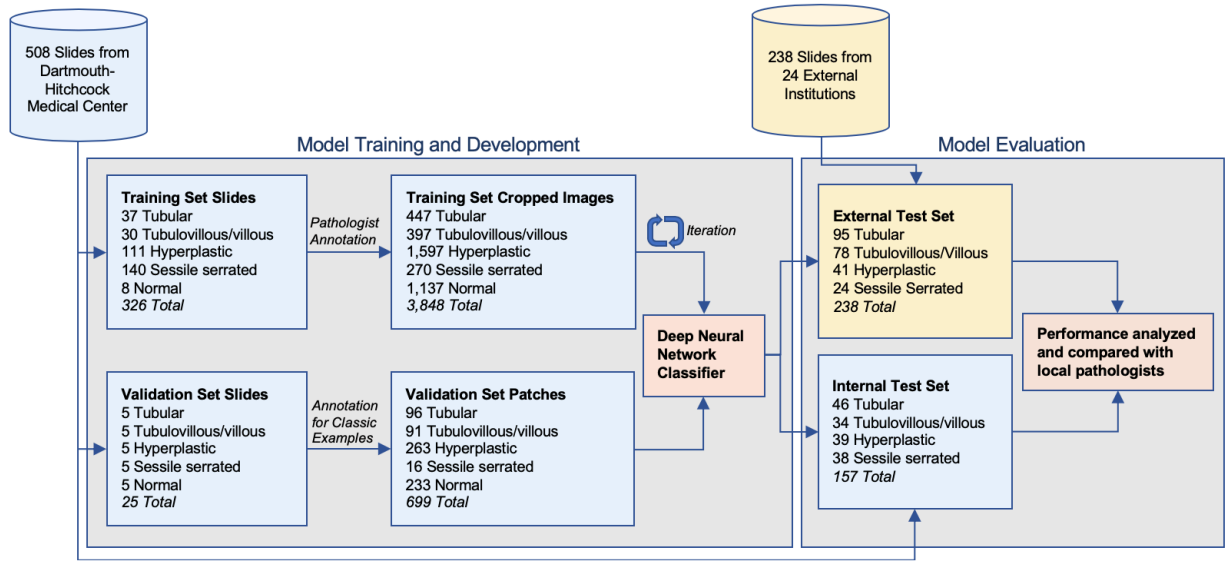

**Figure 1.** Data flow diagram for our study. We trained our model on an internal training and validation set and then evaluated it on internal and external test sets with multi-pathologist gold-standard diagnoses. WSI denotes whole-slide image. Annotated regions of interest in the training set varied in length and width, whereas patches in the validation set were of fixed size and represented classic examples of each polyp type.

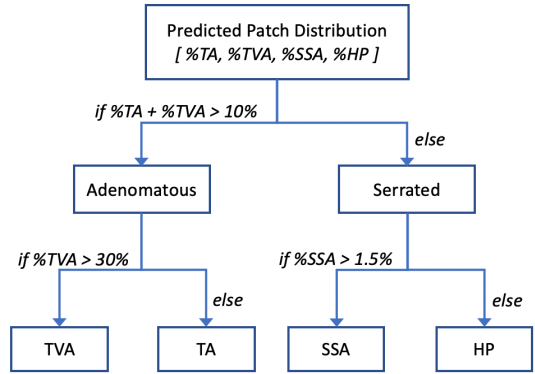

**Figure 2.** Diagram demonstrating thresholds for hierarchical classification of whole-slide images based on patch-level predictions.

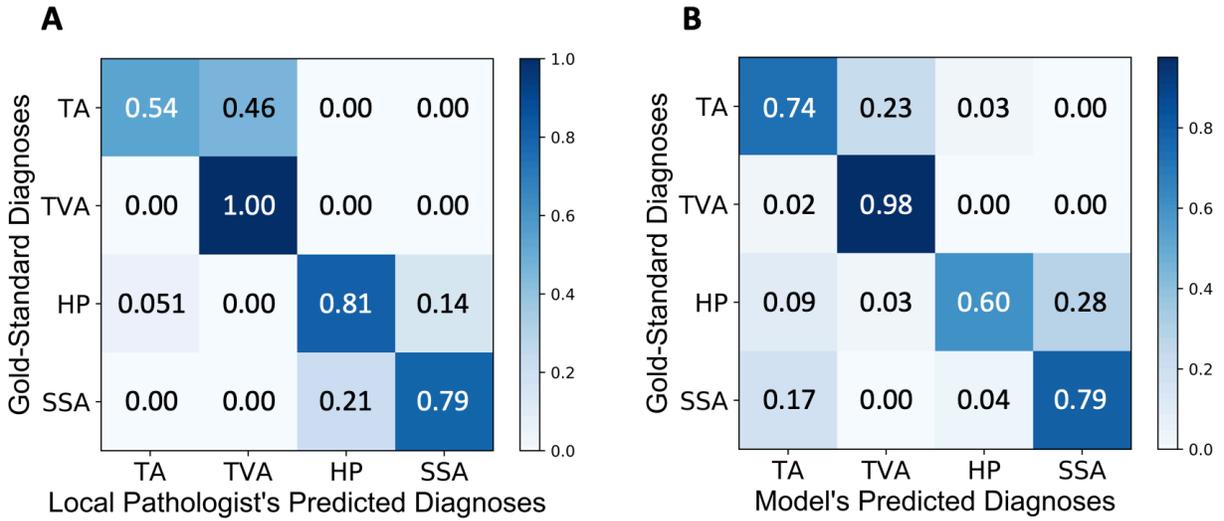

**Figure 3.** Confusion matrices for (**A**) local pathologists' diagnoses given at the point-of-care and (**B**) the model's predicted diagnoses in comparison with multi-pathologist gold-standard diagnoses for the external test set. Each cell in the confusion matrix is the agreement between multi-pathologist gold standard labels and local pathologists' or our model's diagnoses. TA: tubular adenoma; TVA: tubulovillous/villous adenoma; HP: hyperplastic polyp; SSA: sessile serrated adenoma.

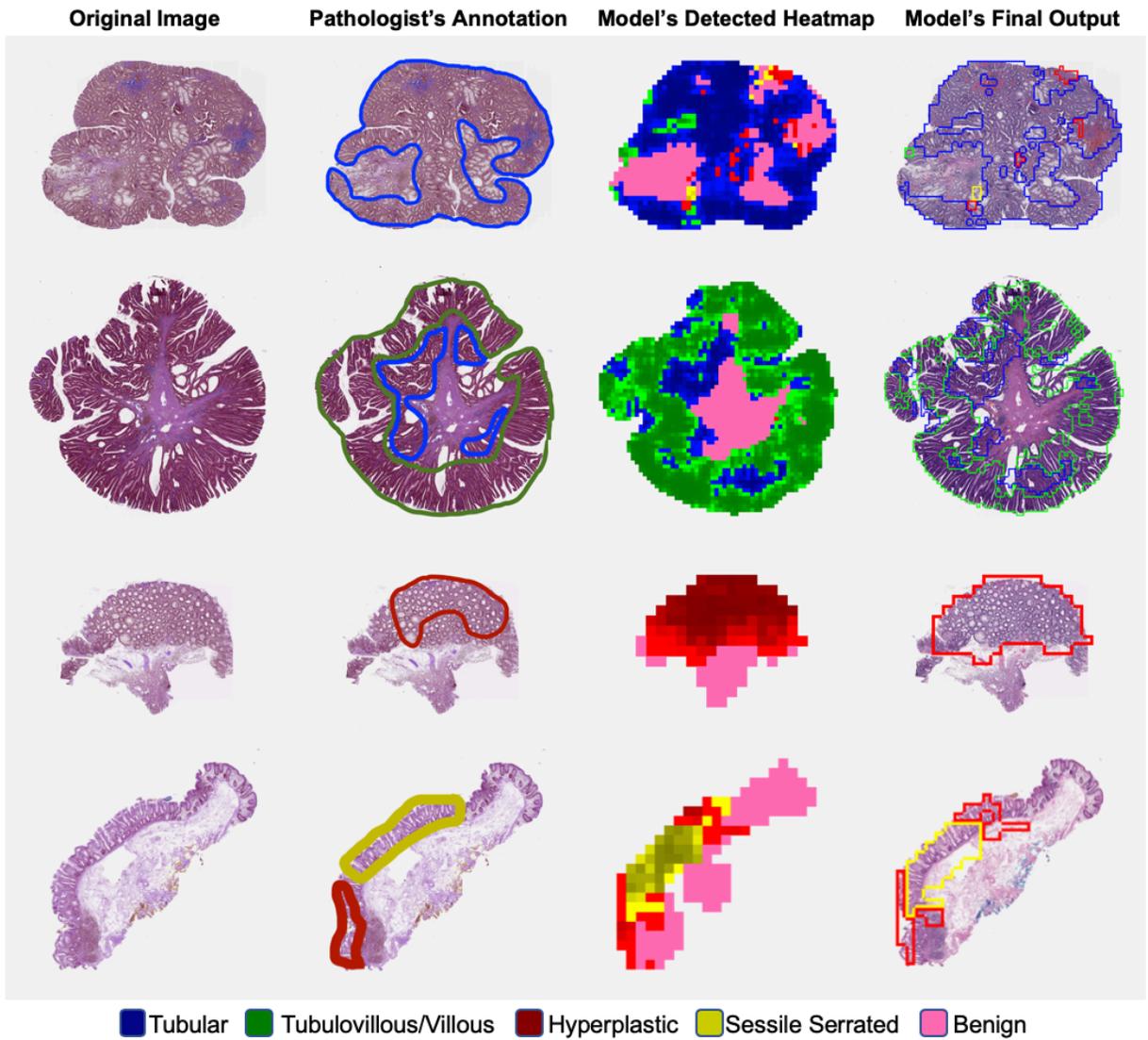

**Figure 4.** Visualization of the classifications of the deep neural network model. The first column shows the original image, and the second column shows pathologist annotations of polyps. The third column depicts the model's detected heatmap, where higher confidence predictions are shown in darker color. In the fourth column, we show the model's final output, which highlights precancerous lesions that can potentially be used to aid pathologists in clinical practice.

# SUPPLEMENTARY MATERIALS

**Supplementary Table 1.** Colorectal polyp slide class distribution for our multi-institutional external test set grouped by pathology laboratory institutional affiliation type and state. TA: tubular adenoma; TVA: tubulovillous/villous adenoma; HP: hyperplastic polyp; SSA: sessile serrated adenoma.

| Pathology Lab Affiliation Type | State | Pathologist Consensus Diagnosis | | | | |
| --- | --- | --- | --- | --- | --- | --- |
| | | TA | TVA | HP | SSA | Total Slides |
| University hospital | Georgia | 3 | 0 | 25 | 2 | 30 |
| | Colorado | 0 | 4 | 7 | 0 | 11 |
| | California | 3 | 0 | 0 | 0 | 3 |
| | Texas | 0 | 0 | 1 | 0 | 1 |
| Veteran's hospital | Georgia | 17 | 1 | 0 | 0 | 18 |
| | Minnesota | 6 | 3 | 1 | 0 | 10 |
| | Colorado | 1 | 0 | 4 | 0 | 5 |
| Metropolitan/regional hospital | Iowa | 6 | 3 | 3 | 2 | 14 |
| | Ohio | 4 | 1 | 6 | 5 | 16 |
| | Minnesota | 0 | 0 | 2 | 5 | 7 |
| | South Carolina | 3 | 1 | 3 | 1 | 8 |
| | New Hampshire | 4 | 0 | 1 | 0 | 5 |
| | South Carolina | 2 | 0 | 0 | 0 | 2 |
| | Minnesota | 0 | 0 | 0 | 1 | 1 |
| | California | 2 | 0 | 0 | 0 | 2 |
| | Iowa | 1 | 0 | 1 | 0 | 2 |
| Freestanding | Iowa | 15 | 3 | 4 | 1 | 23 |
| | Iowa | 10 | 9 | 5 | 0 | 24 |
| | Iowa | 2 | 0 | 3 | 0 | 5 |
| | Colorado | 0 | 0 | 1 | 0 | 1 |
| | New Hampshire | 1 | 0 | 2 | 0 | 3 |
| | Colorado | 0 | 0 | 3 | 0 | 3 |
| | Florida | 0 | 0 | 1 | 0 | 1 |
| Specialty clinic/practice | Minnesota | 15 | 16 | 5 | 7 | 43 |
| Combined | | 95 | 41 | 78 | 24 | 238 |

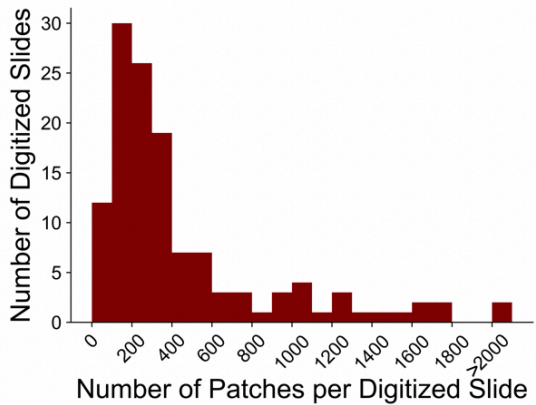
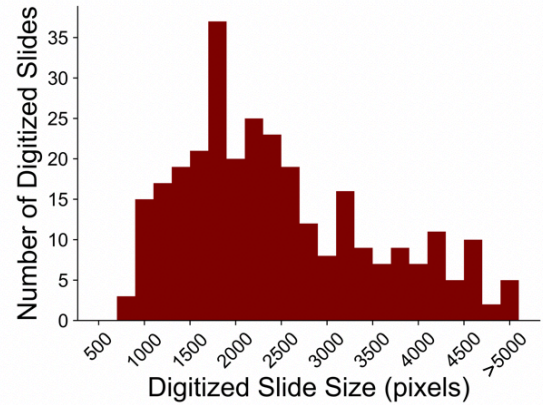
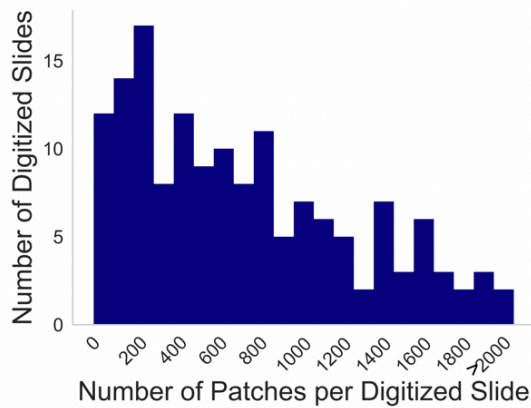
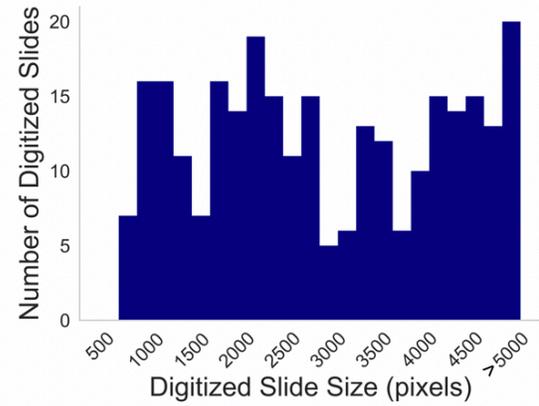

**Supplementary Figure 1.** Number of patches per digitized slide and slide size (in pixels) for **(A)** the internal test set and **(B)** the multi-institutional external test set. Patches are fixed-size areas of tissue obtained by sliding a window over the entire image. Size of digitized slides reflects the area of the tissue after removing the background.

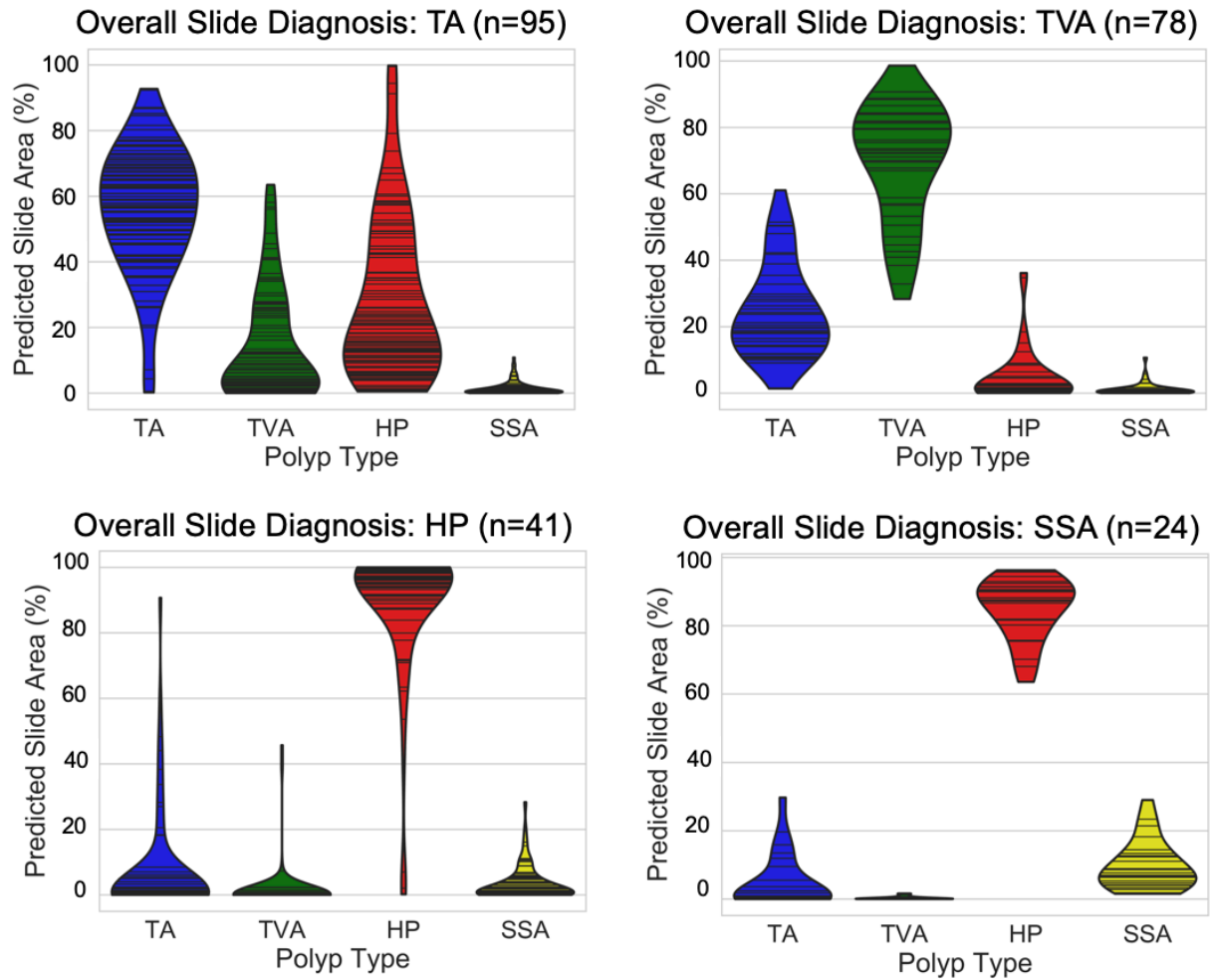

**Supplementary Figure 2.** Violin plots showing predicted percentage areas (based on number of patches) for each polyp type on whole-slide images. TA: tubular adenoma; TVA: tubulovillous/villous adenoma; HP: hyperplastic polyp; SSA: sessile serrated adenoma. Subjective inspection by our study pathologists confirm that each plot reflects the expected histology distribution. For whole slides that were diagnosed as TA and TVA, our model detected significant areas of TA and TVA, reflecting the morphological similarity of the two polyp types. Our model also detected large areas of HP in whole slides diagnosed as TA, which is expected

since all polypoid lesions in TAs are exposed to elevated mechanical forces and therefore show hyperplastic features at their peripheries. While the model detected mostly HP areas in whole slides diagnosed as hyperplastic polyps, it did find some small areas of SSAs, possibly because larger HPs with deep, dilated crypts and serrated epithelium can appear similar to SSAs. Finally, for whole slides that were diagnosed as SSAs, it makes sense that HPs comprised the largest area, since all SSAs have significant morphological overlap with hyperplastic polyps and may contain only a few classic, broad-based, dilated crypts with heavy serration.